\documentclass[10pt,prl,aps,twocolumn,floatfix,nofootinbib]{revtex4-2}
\usepackage{graphicx}
\usepackage{amssymb}
\usepackage{amsmath}
\usepackage{xcolor}
\newcommand{\Co}{C^O}
\newcommand{\Cp}{C^\Phi}
\newcommand{\CC}{{\cal C}}
\newcommand{\CCp}{{\cal C}^\Phi}

\begin{document}

\title{Disentangling Multiple Scaling Dimensions via Operator Covariance \\ in Critical Quantum Spin Systems}

\author{Zhi Lin}
\affiliation{Department of Physics, Boston University, 590 Commonwealth Avenue, Boston, Massachusetts 02215}

\author{Anders W. Sandvik}
\affiliation{Department of Physics, Boston University, 590 Commonwealth Avenue, Boston, Massachusetts 02215}

\begin{abstract}
We present a general proof of different power-law decays versus distance of the eigenvalues of two-point covariance matrices in critical lattice
quantum models, allowing extraction of multiple scaling dimensions. We demonstrate the practical utility of this method
in the context of a spin-$1/2$ chain with critical ground state described by the SU($2$)$_1$ Wess--Zumino--Witten field theory.
Using quantum Monte Carlo simulations to compute covariance matrices in different symmetry sectors, we resolve the expected relevant and
marginal primary operators as well as some descendants. The covariance method can be integrated into many existing numerical schemes and promises
to significantly enhance computational studies of quantum criticality.
\end{abstract}

\date{\today}

\maketitle

Critical points are characterized by scaling exponents governing long-distance correlation functions. The conventional exponents are related
to scaling dimensions of relevant perturbations that destabilize the critical point, while irrelevant perturbations cause corrections to scaling.
Determining scaling dimensions is a key task for computational studies of classical and quantum criticality, and there is a good
general understanding of how to analyze correlation functions or, alternatively, to obtain exponents from finite-size scaling
of observables slightly off criticality \cite{Fisher72,Barber83,Pelissetto02,Sandvik10}. There are still challenges, however, related to
limitations of the system size in quantum Monte Carlo (QMC) simulations and difficult data analysis when the scaling corrections are large.

Multicritical points pose particular challenges due to the presence of more than one relevant scaling dimension in correlation functions, necessitating
complex data fits that are highly sensitive to statistical errors and often deliver reliable results only for the leading decay power. Many intriguing
open problems involve potential multicriticality, e.g., the deconfined quantum critical point \cite{Senthil04,Takahashi24,Chester24,Li26} and the
point separating a deconfined phase and a self-dual first-order line in lattice gauge models \cite{Somoza21}. Techniques capable of resolving
more than one relevant scaling dimension could therefore have significant impact.

We here build on the recent proposal \cite{Sandvik26} to use distance-dependent covariance matrices
\begin{equation}\label{cdef}
 \Co_{ab}(r)=\langle O^\dagger_a(0)O_b(r)\rangle,~~ a,b \in \{1,\ldots,N\},
\end{equation}
of lattice operators, where a nonzero expectation value $\langle O^\dagger_a\rangle\langle O_b\rangle$ has been implicitly subtracted off.
It was argued in Ref.~\cite{Sandvik26} that the eigenvalues $\lambda_k(r)$ of the $N\times N$ matrices $\Co(r)$ decay with different exponents corresponding
to scaling dimensions of the continuum field operators contained in the set $\{O\}$,
\begin{equation}\label{lambdan}
\lambda_k(r) \sim r^{-2\Delta_k}, 
\end{equation}
with the accessible $\Delta_k$ restricted by the symmetries of the operators used. In contrast, the conventional correlators 
$\Co_{ii}(r)$ mix scaling dimensions and only the slowest-decaying component can be reliably extracted.

Arguments were put forward for the separation of scaling dimensions in the asymptotic eigenvalues, but a rigorous proof
was not presented \cite{Sandvik26}. A proof does exist for the separation of exponential decays via (generalized) covariance eigenvalues $\lambda_k(\tau)$ in
imaginary time $\tau$ in the case of discrete energy levels $k$ \cite{Lucher90}. This approach is often used in lattice calculations in particle
physics to extract masses \cite{Peardon97,Peardon99,Alpha09,Liu12} and can also in principle be applied to discrete spectra of finite lattice systems
with emergent conformal field theory (CFT) description. For such systems in the correct geometry (corresponding to radial quantization), low-energy levels
of the Hamiltonian are directly related to scaling dimensions \cite{Cardy84,Cardy86}. However, in this context we are only aware of work on individual
conventional correlation functions \cite{Milsted17,You18,Hoffmann23,Ayyar23}.

Here we focus on real-space covariance but expect the method to also resolve multiple power-law decays in imaginary time. We will present a general proof
of Eq.~(\ref{lambdan}) for a system described by a Euclidean Lorentz-invariant theory, i.e., with dynamic exponent $z=1$. While the method may also
work if $z \not= 1$, our proof relies on a reflection positivity that would have to be confirmed. As a test case, we consider a quantum-critical
antiferromagnetic spin-$1/2$ chain described by the Wess--Zumino--Witten (WZW) SU($2$)$_1$ CFT.
The continuum operator content of the spin operators
is well understood \cite{Affleck87} and is more complicated than the Ising case considered previously \cite{Sandvik26}.

{\it Eigenvalue decays}.---Critical
fluctuations in quantum lattice models arise from gapless points $q=Q$ in momentum space. We will here work with the case of
$Q=0$ and $Q=\pi$ in one space dimension, but generalization to other gapless points is possible. The continuum fields $\Phi_\mu(x)$
are smooth and the lattice momentum is introduced when expanding operators,
\begin{equation}\label{OPhiexp}
O_a(x)=\sum_\mu z_{a\mu}{\rm e}^{iQ_\mu x}\Phi_\mu(x), 
\end{equation}
where $x$ is the integer lattice coordinate and $z_{a\mu}$ the properly defined overlaps. Our proof
of Eq.~(\ref{lambdan}) relies on general field-theory facts that in combination with theorems for Hermitian matrices imply positive-definite
covariance matrices in cases where lattice operators do not mix different momenta (e.g., with a single gapless point); otherwise our proof still
holds for subsets of distances $r$, depending on how the momenta are mixed. 

Before discussing the lattice matrix $C^O(r)$, we first consider the continuum scaling fields themselves, for which positive semidefiniteness
follows from reflection positivity. At a scale-invariant fixed point, let $\Theta$ denote the spatial reflection $x\to -x$ about the midpoint
between two $r=2x$ separated insertion points combined with Hermitian conjugation. Correlators defined with this antiunitary operator take the
form \cite{Yellowbook,Rychkov17}
\begin{equation}\label{Cftdef}
\CCp_{\mu\nu}(r) = \langle \Theta\Phi_\mu(r/2)\Phi_\nu(r/2)\rangle
= G_{\mu\nu}r^{-(\Delta_\mu+\Delta_\nu)}
\end{equation}
with an $r$-independent matrix $G$. In a unitary Lorentz-invariant theory, $\Theta$ is the famous Osterwalder--Schrader (OS) time reflection
(under which the theory is reflection positive) rotated into a spatial reflection \cite{Osterwalder73,Osterwalder75,Glimm87}. Since
$\langle \Theta FF\rangle$ is the squared norm of a state $|F\rangle$, any $F_c(x)=\sum_\mu c_\mu \Phi_\mu(x)$ satisfies
$\langle\Theta F_c(x)F_c(x)\rangle \ge 0$, and, thus, for any vector $c=(c_1,c_2,\ldots)$
\begin{equation}
\sum_{\mu\nu} c^*_\mu c_\nu\CCp_{\mu\nu}(r) = c^\dagger R(r)GR(r)c \ge 0,
\end{equation}
where $R(r)$ is the diagonal matrix
\begin{equation}
R_{\mu\nu}(r)=\delta_{\mu\nu}r^{-\Delta_\mu}.
\end{equation}
Since $R(r)$ is invertible for $r>0$, the vector $y=R(r)c$ is as arbitrary as $c$, so $y^\dagger G y\ge 0$ for any $y$ and $G$ is therefore Hermitian and
positive semidefinite. In a proper theory with null operators removed, $y^\dagger G y\not= 0$ for $y \not=0$, $G$ is positive definite and can be written as
\begin{equation}\label{gkk}
G = K K^\dagger
\end{equation}
with a nonsingular matrix $K$.

We can impose parity $p_\mu=\pm 1$ under $x \to -x$ so that $\Theta \Phi_\mu(r/2) = p_\mu t_\mu \Phi_\mu^\dagger(-r/2)$, where the antiunitarity of
$\Theta$ implies that also the time parity $t_\mu = \pm 1$ of $\Phi_\mu$ appears. We can then define the ordinary correlation function
$\Cp_{\mu\nu}(r)=p_\mu t_\mu \CCp_{\mu\nu}(r)$ as
\begin{equation}\label{Cftdef2}
\Cp_{\mu\nu}(r) = \langle \Phi^\dagger_\mu(-\hbox{$\frac{r}{2}$})\Phi_\nu(\hbox{$\frac{r}{2}$})\rangle
 =  p_\mu t_\mu G_{\mu\nu}r^{-(\Delta_\mu+\Delta_\nu)},
\end{equation}
with the same $G$ as before. With common $(p_\mu,t_\mu)$ for all $\mu$, the common sign is unimportant in scaling. In one space dimension the reflection
$x \to -x$ is unique and a field of Lorentz spin $s_\mu$ has $p_\mu t_\mu =(-1)^{s_\mu}$. In higher dimensions, the relevant reflection is with respect to the plane
perpendicular to the separation line and a field of given $s_\mu$ need not have a definite parity under this reflection, but fields of definite $p_\mu$
can still be chosen.

Like the fields, the lattice operators $\{O\}$ can also be constructed with fixed parity with respect to a chosen mirror plane. To compare with
Eq.~(\ref{Cftdef}), the reflection operator should be of the same antiunitary form as $\Theta$, and we therefore define
\begin{equation}
\Theta_{\rm lat}={\cal R T} = {\cal R_{\rm c} T}T_{-r},
\end{equation}
where ${\cal R}$ reflects all lattice operators $x \to -x$, ${\cal R}_{\rm c}$ similarly reflects a single lattice operator about its midpoint
(which can be a site or a bond), $T_r$ translates by $r$ lattice constants, and ${\cal T}$ is time reversal. Thus, acting on a spin operator,
$\Theta_{\rm lat}{\bf S}_i=-{\bf S}_{i'}$, where site $i'$ is $i$ under $x \to -x$. We choose all $O_a$ to have the same parity $p_a$ and time-reversal
eigenvalue $t_a=\pm 1$, so that $\Theta_{\rm lat}O_a(x)=p_at_aO^\dagger_a(-x)$, and then define
\begin{eqnarray}
\CC_{ab}(r) & = & \langle \Theta_{\rm lat} O_a(\hbox{$\frac{r}{2}$})O_b(\hbox{$\frac{r}{2}$})\rangle \nonumber \\
    & = & p_at_a \langle O^\dagger_a(-\hbox{$\frac{r}{2}$})O_b(\hbox{$\frac{r}{2}$})\rangle. \label{Cposdef}
\end{eqnarray}
In practice, we work with Hermitian operators, $O^\dagger_a=O_a$, but here use notation conforming with Eq.~(\ref{Cftdef2}).

A cutoff at $M$ fields can be imposed in proofs (and finally $M\to \infty$). In Appendix A, we first use Eq.~(\ref{Cftdef}) and theorems for positive-definite
Hermitian matrices to prove that the eigenvalues $\lambda_k(r)$ of an $M\times M$ covariance matrix $\CCp(r)$ are bounded according to
\begin{equation}\label{bounds}
a_kr^{-2\Delta_k} \le \lambda_k \le b_k r^{-2\Delta_k},
\end{equation}
with $r$-independent $a_k \le b_k$ and with the eigenvalues (scaling dimensions) decreasing (increasing) with $k$. Given this result, in Appendix A we
further prove that the eigenvalues of $N \times N$ lattice covariance matrices $C^O(r)$ expressed with $M$ scaling fields as in Eq.~(\ref{OPhiexp}) can also
be bounded in a similar way. Thus, Eq.~(\ref{lambdan}) is proven. In the presence of scaling corrections (beyond those from field mixing) due to irrelevant
perturbations, the eigenvalues acquire corrections, $\lambda_k \to \lambda_k(1+\sum_j\epsilon_{kj} r^{-\omega_j})$, but there is no change in the decay exponents.

{\it Mixing momentum components}.---It is
well known that critical one-dimensional SU(2)-symmetric spin models, such as the Heisenberg chain, have gapless momentum points at $Q=0$ and
$Q=\pi$ and the way the lattice operators overlap with the continuum fields is then partially affected by minus signs in Eq.~(\ref{OPhiexp})
originating from the $Q=\pi$ operator content. To determine the sign structure, we have to consider how the signs ${\rm e}^{iQ_\mu x}$ of the $Q=\pi$
components in Eq.~(\ref{OPhiexp}) transform when an operator defined on $w$ sites $i,i+1,\ldots,i+w-1$ is reflected $x \to -x$ under $\Theta_{\rm lat}$.
The spatial reflection is either bond- or site-centered, depending on the separation $r$ as well as the width $w$ of the operator. By comparing the
staggered signs $(-1)^x$ of the $\Theta_{\rm lat}$-reflected and original sites of the operator cell centered at $x=r/2$, it can be easily seen that
the overlap with the $Q=\pi$ components must come with a sign $(-1)^{r+\kappa}$, with $\kappa=w \bmod 2$. Thus, the lattice covariance matrix
takes the form
\begin{equation}\label{CCsplit}
\CC(r) = \CC_0(r) + (-1)^{r+\kappa}\CC_\pi(r),
\end{equation}
where the individual covariance matrix elements also involve the overall parity $p_at_a$ as in Eq.~(\ref{Cposdef}) and it is understood that all operators
are defined with either even or odd $w$ ($\kappa=0,1$) and have definite parity with respect to reflection about the cell center (a bond or a site for
$\kappa=0$ and $\kappa=1$, respectively). It is not possible to construct lattice operators overlapping only with the $Q=0$ or $Q=\pi$ fields---which is
trivial in the sense that a finite lattice operator does not have a momentum. To ensure positive definiteness of a mixed $\CC(r)$, the distances $r$
can be restricted to even or odd, depending on $w$, so that both terms have the same sign in Eq.~(\ref{CCsplit}).

The mapping between spin operators and the fields of the WZW level-$1$ theory is known \cite{Affleck87,Yellowbook}:
\begin{equation}\label{sop}
\mathbf{S}_i \sim \mathbf{J}_L(x_i) + \mathbf{J}_R(x_i) + a(-1)^i\mathbf{n}(x_i) +\cdots,
\end{equation}
where $\mathbf{n}$ ($\Delta_n =1/2$, $s=0$) is the N\'eel order parameter and
$\mathbf{J}=\mathbf{J}_L+\mathbf{J}_R$ ($\Delta_J=1$, $s=\pm 1$) is the uniform spin density. 
The bond operator at $\bar x_i=x_i+1/2$ is 
\begin{eqnarray}\label{ssop}
&& \mathbf{S}_i\cdot \mathbf{S}_{i+1} \sim
b_0 \mathbb{I} +b_1(-1)^i\epsilon(\bar x_i) + b_2\mathbf{J}_L(\bar x_i)\cdot \mathbf{J}_R(\bar x_i)~ \nonumber \\
&& ~~~~~~~~~~~~~~~~~~~~ + b_3 [T_L(\bar x_i)+T_R(\bar x_i)] + \cdots ,
\end{eqnarray}
where $\epsilon$ ($\Delta_\epsilon =1/2$, $s=0$) is the dimerization field, $T_L(x)+T_R(x)$ is the energy density ($\Delta_T=2$, $s=\pm 2$),
and $\mathbf{J}_L\cdot \mathbf{J}_R$ ($\Delta_{JJ}=2$, $s=0$) is the marginal operator that drives the dimerization transition. The
constant $\mathbb{I}$ is not present in connected correlators. 

The fields in Eq.~(\ref{sop}) and Eq.~(\ref{ssop}) have $t=-1$ and $t=+1$, respectively, and the spatial parity is $p=t(-1)^s$ for
bond-centered reflections. For a site-centered cell, the bond-centered reflection must be combined with a one-site translation, which gives an extra
factor $-1$ for the staggered fields, corresponding to ${\rm e}^{i\kappa Q_\mu}$. Note that the double degeneracies for $s\not=0$ are invisible for
operators constructed at equal time, where a real combination of the two cases is realized.

The operators $O_a$ combine spin operators within a cell of $w$ sites, adding different motifs to achieve even or odd spatial parity. Such combinations
can cancel out the leading $Q=0$ or $Q=\pi$ fields but also generate additional operator content of the extended operators,
e.g., $O_+=\mathbf{S}_i+\mathbf{S}_{i+\delta}$ with odd separation $\delta$ cancels
${\bf n}$ but generates $\partial_x {\bf n}$, while for even $\delta$ the ${\bf n}$ field is amplified. Conversely, $O_-=\mathbf{S}_i-\mathbf{S}_{i+\delta}$
for odd $\delta$ cancels the uniform terms but generates their derivatives. Note, however, that canceling a leading field of given $Q_\mu$ does not
imply that the operator excludes an entire momentum sector. For an operator $O_a$ to overlap with a field $\Phi_\mu$ the selection rules
$t_a=t_\mu$ and $p_a = {\rm e}^{i\kappa Q_\mu}p_\mu$ must be satisfied. For given $p_a$, the latter can be satisfied by $Q_\mu =0$ as well as $Q_\mu =\pi$ fields.

In general, when exploring models for which the continuum operator content of lattice operators is not known, it will not be a priori clear
what scaling dimensions appear in a given lattice operator, but information gained from covariance matrices with different classes of
operators can help to establish the mapping. The above selection rules can be easily generalized to antiferromagnets in higher
dimensions, e.g., in the case of deconfined quantum criticality, where the gapless points are $Q=(0,0)$, $Q=(\pi,0)$, $Q=(0,\pi)$, and $Q=(\pi,\pi)$ 
\cite{Senthil04,Suwa15,Ma18,Yang25}.

{\it Numerical results}.---The Heisenberg spin-$1/2$ chain is described the WZW CFT \cite{Affleck87,Eggert96}, but its
standard definition with only nearest-neighbor couplings $J{\bf S}_i \cdot {\bf S}_{i+1}$ contains a marginally irrelevant perturbation causing logarithmic
scaling corrections. The coupling to the marginal operator can be tuned to $0$, the point beyond which it changes sign and becomes marginally relevant,
causing spontaneous dimerization. Many different extensions of the Heisenberg chain allowing such tuning are known \cite{Nielsen11}, e.g., including
second-neighbor couplings $J'{\bf S}_i \cdot {\bf S}_{i+2}$ \cite{Eggert96} or spin-phonon couplings \cite{Suwa16}. For large-scale QMC
simulations, a convenient choice is the $J$-$Q$ chain defined by
the Hamiltonian
\begin{equation}
H = - J\sum_{i=1}^L P_{i,i+1} - Q\sum_{i=1}^L P_{i,i+1}P_{i+2,i+3},
\end{equation}
where $P_{i,j}=1/4-{\bf S}_i \cdot {\bf S}_{j}$ with periodic boundary conditions. The dimerization transition is at $(Q/J)_c=0.84831$
and power-law decaying correlations (i.e., the leading powers governed by $\Delta_n$ and $\Delta_\epsilon$) without signs of logarithmic corrections (which are
very prominent in the Heisenberg chain, $Q=0$) have been observed \cite{Sanyal11,Tang11,Patil18}.

\begin{table}[t]
\caption{Operators used here in covariance matrices computed with QMC simulations. The subscripts of the spin operators
correspond to consecutive labels in a cell of $w$ sites. Covariance matrices are constructed for the groups of operators with
$t$ (time) and $p$ (space) parity.}
\begin{tabular}{|l||l|l|l|l|}
\toprule  ~ ID ~ &  ~ Operator & ~ $w$ ~ & ~ $t$ ~ & ~ $p$ ~\cr
\hline
~ $O_1$ ~ & ~ $\mathbf{S}_2 \cdot \mathbf{S}_3$ ~ & ~ $4$ ~ & ~ $+$ ~ & ~ $+$ ~ \cr \hline
~ $O_2$ ~ & ~ $\mathbf{S}_1 \cdot \mathbf{S}_4$ ~ & ~ $4$ ~ & ~ $+$ ~ & ~ $+$ ~ \cr \hline
~ $O_3$ ~ & ~ $(\mathbf{S}_1 \cdot \mathbf{S}_2 + \mathbf{S}_3 \cdot \mathbf{S}_4)/2$ ~ & ~ $4$ ~ & ~ $+$ ~ & ~ $+$ ~ \cr \hline
~ $O_4$ ~ & ~ $(\mathbf{S}_1 \cdot \mathbf{S}_3 + \mathbf{S}_2 \cdot \mathbf{S}_4)/2$ ~ & ~ $4$ ~ & ~ $+$ ~ & ~ $+$ ~ \cr \hline
~ $O_5$ ~ & ~ $(S^z_1 + S^z_2 + S^z_3 + S^z_4)/4$ ~ & ~ $4$ ~ & ~ $-$ ~ & ~ $+$ ~ \cr \hline
~ $O_6$ ~ & ~ $(S^z_1 + S^z_4)(S^z_2S^z_3)$ ~ & ~ $4$ ~ & ~ $-$ ~ & ~ $+$ ~ \cr \hline
~ $O_7$ ~ & ~ $(S^z_2 + S^z_3)(S^z_1S^z_4)$ ~ & ~ $4$ ~ & ~ $-$ ~ & ~ $+$ ~ \cr \hline
~ $O_8$ ~ & ~ $(\mathbf{S}_1 \cdot \mathbf{S}_2 - \mathbf{S}_4 \cdot \mathbf{S}_5)/2$ ~ & ~ $5$ ~ & ~ $+$ ~ & ~ $-$ ~ \cr \hline
~ $O_9$ ~ & ~ $(\mathbf{S}_2 \cdot \mathbf{S}_3 - \mathbf{S}_3 \cdot \mathbf{S}_4)/2$ ~ & ~ $5$ ~ & ~ $+$ ~ & ~ $-$ ~ \cr \hline
~ $O_{10}$ ~ & ~ $(\mathbf{S}_1 \cdot \mathbf{S}_4 - \mathbf{S}_2 \cdot \mathbf{S}_5)/2$ ~ & ~ $5$ ~ & ~ $+$ ~ & ~ $-$ ~ 
\cr \botrule
\end{tabular}
\label{otable}
\end{table}

\begin{figure}[t]
\includegraphics[width=80mm]{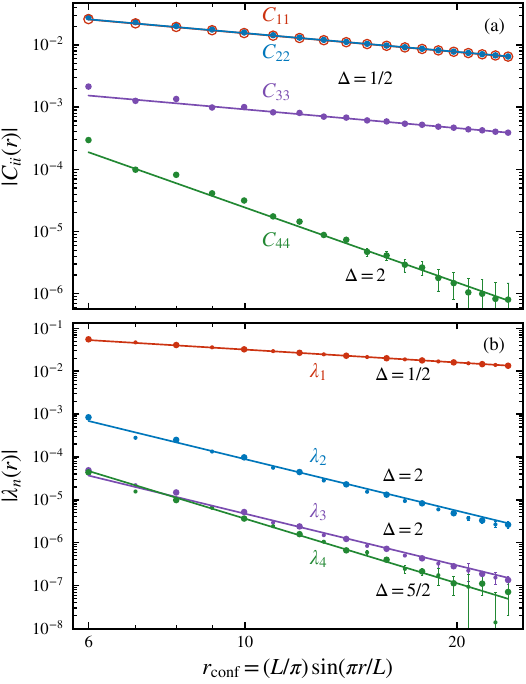}
\caption{Individual correlation functions (a) and covariance eigenvalues (b) computed with the $w=4$ time-reversal even, parity-even operators
in Table~\ref{otable}. The signs are even-odd oscillating and the absolute values are therefore shown. Decay powers corresponding to expected scaling
dimensions are indicated by the lines. In (b), the smaller symbols are for odd $r$, for which positive definiteness of the covariance matrix cannot
be guaranteed because of mixed signs in Eq.~(\ref{CCsplit}).}
\label{fig1}
\end{figure}

\begin{figure}[t]
\includegraphics[width=80mm]{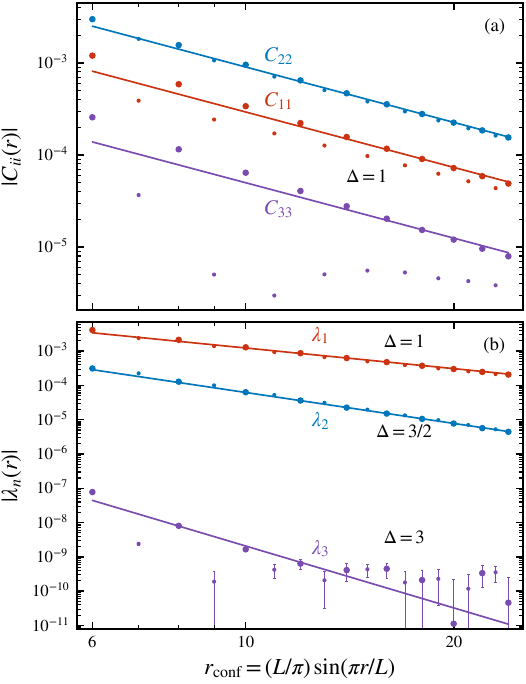}
\caption{Individual correlation functions (a) and covariance eigenvalues (b) computed with the $w=4$ time-reversal odd, parity-even operators
in Table~\ref{otable}. Decay powers corresponding to expected scaling dimensions are indicated by the lines. Smaller and larger circles have
the same meaning as in Fig.~\ref{fig1}.}
\label{fig2}
\end{figure}

We construct three covariance matrices based on the operators in Table~\ref{otable}, grouped by their cell size $w$ and
parity $t$, $p$. QMC calculations were performed using the ground-state projector method in the valence-bond basis \cite{Sandvik10b}, with the
system size fixed at $L=256$. In Figs.~\ref{fig1} ($w=4$, $t=+1$, $p=+1$) and \ref{fig2} ($w=4$, $t=-1$, $p=+1$), we graph results for the
conventional correlation functions $\Co_{ii}(r)$ and the covariance eigenvalues in the (a) and (b) panels, respectively. We show
results only for $r$ up to where the fastest decaying eigenvalue becomes noisy, but the other eigenvalues remain well behaved also for much
larger $r$. We plot the results versus the conformal (chord) distance $r_{\rm conf}$, which is not much different from $r$ for $r \ll L/2$.

The operators $O_1$, $O_2$, and $O_3$ mix uniform and staggered fields, while $O_4$ does not overlap with the staggered $\epsilon$ field
because both ${\bf S}_i \cdot {\bf S}_j$ terms have $i$ and $j$ on the same sublattice. Therefore, $\Co_{11}$, $\Co_{22}$, and $\Co_{33}$
in Fig.~\ref{fig1}(a) decay with the lowest scaling dimension $\Delta_\epsilon=1/2$ in Eq.~(\ref{ssop}) while $\Co_{44}$ is dominated by some
combination of the uniform operators. The eigenvalues reveal all four lowest scaling dimensions, including the doubly degenerate
$\Delta_T=\Delta_{JJ}=2$ (but cannot identify which decay corresponds to which operator) and also $\Delta''_\epsilon=5/2$ from (a mix of)
the $s=0$ and $s=2$ second descendants of $\epsilon$. It is noteworthy that the absolute values of the negative odd-$r$ eigenvalues, for
which the proof of the method formally does not hold, still follow the same trend as the even-$r$ data. When not including $O_4$, both
$\Delta_T=\Delta_{JJ}=2$ are still resolved but with larger errors on the one with the smaller prefactor (not shown here). 

In Fig.~\ref{fig2} we show results with the time-odd operators $O_5$, $O_6$, and $O_7$ based on $S^z_i$ components only and with
four or two operators added so that the N\'eel field cancels out; in $O_6$ and $O_7$ we further multiply by a time-even
product. These operators are $z$-projections of spin $S=1$ and $S=3$ operators, but the latter produce much higher scaling
dimensions and are not visible in practice. The smallest dimension $\Delta_n=1/2$ is indeed absent and the slowest decay instead
corresponds to $\Delta_J=1$ in Eq.~(\ref{sop}). The dimensions $\Delta=3/2$ and $\Delta=3$ correspond to descendants of $n^z$ and $J^z$.
The even- and odd-$r$ eigenvalues again show essentially the same decay forms, with the exception of the $\Delta=3$
case where the deviations for small $r$ may just originate from corrections.

Results for the $(t=+1,p=-1)$ operators $O_8$, $O_9$, and $O_{10}$ are presented in Appendix B and show the presence of the dimerization field
$\epsilon$ and two $\Delta=5/2$ descendants.

{\it Discussion.}---Our proof of disentangling of scaling dimensions relies on positive definiteness, which we ascertained using
field-theory results for systems with dynamic exponent $z=1$. Positive definiteness by itself is not sufficient, since the form Eq.~(\ref{Cftdef})
with $r$-independent $G$ (at least for large $r$) is a prerequisite. This may, however, not strictly require Lorentz invariance even
though we used it to rotate the OS reflection to a spatial plane; thus, the proof can likely be extended to $z\not=1$ by demonstrating spatial
reflection positivity without reference to the time dimension. Moreover, working with time correlations, the OS reflection is not rotated
and the proof goes through with modified scaling dimensions $\Delta_k \to \Delta_k/z$.

In practice, when a proof formally is not applicable, support for the correctness of the method can still be gathered empirically, by finding a
positive-definite covariance matrix with large-$r$ separated power laws of the eigenvalues and convergent eigenvectors. It is then very likely that
observed power-law decays really reflect scaling dimensions of a low-energy theory.

The covariance method can also be formulated as a generalized eigenvalue problem, in analogy with Ref.~\cite{Lucher90},
which we discuss in Appendix C. The results do not change appreciably. Covariance could also possibly be exploited with
``causation'' \cite{Wichmann26}.

{\it Note added.}---At the final stages of this work we became aware of complementary work by Chen and Kaul \cite{Chen26}. They presented a
different proof of the method in the classical setting and also tested eigenvalue separation with the
exact solution of the two-dimensional Ising model. Moreover, they analyzed correlation amplitudes, thus further enhancing the utility of
covariance matrices.

\begin{acknowledgments}
{\it Acknowledgments.}---This research was supported by the Simons Foundation under Grant No.~511064 and by a Fellowship from the
John Simon Guggenheim Memorial Foundation (A.W.S.). The numerical calculations were carried out on the Shared Computing Cluster managed by
Boston University's Research Computing Services.
\end{acknowledgments}

\appendix

\subsection*{End Matter}

{\it Appendix A: Eigenvalue decays.}---Building on the demonstration of positive-definite Hermitian matrix $G$ in Eq.~(\ref{Cftdef}), we here
complete the proof of Eq.~(\ref{lambdan}). Before analyzing the covariance matrix of lattice correlations, we consider the case of the fields
$\Phi_a$, i.e., the problem of diagonalizing the right-hand side of Eq.~(\ref{Cftdef}).
For simplicity we impose $\Delta_1 \le \Delta_2 \le \cdots \le \Delta_M$ with a cutoff $M$ (and $M \to \infty$ finally). This step will show that
the off-diagonal elements of $G$ change the amplitudes of the asymptotic decays $r^{-2\Delta_k}$ but the exponents remain
for $r \to \infty$. We then generalize to $N$ lattice operators overlapping with $M$ fields.

We rely on two well-known properties of eigenvalues $\lambda_k$ of Hermitian matrices $A,B$ \cite{Horn13}: {\it (i) Ordering theorem}:
if $c^\dagger A c \le c^\dagger B c$ $\forall$ vectors $c$, then $\lambda_k(A) \le \lambda_k(B)$ $\forall ~k$. {\it (ii) Ostrowski's theorem}: With an
invertible matrix $S$, $\lambda_k(SAS^\dagger) = \gamma_k\lambda_k(A)$, where $\lambda_{\rm min}(SS^\dagger) \le \gamma_k \le \lambda_{\rm max}(SS^\dagger)$,
with min (max) referring to the smallest (largest) eigenvalue.

To analyze the field matrix $\CCp = R(r)GR(r)$, we first apply a Gram--Schmidt procedure, $G=L\Gamma L^\dagger$, where $L$ is lower triangular
with $L_{ii}=1$ and $\Gamma$ is diagonal, $\Gamma={\rm diag}(g_1,\ldots,g_M)$ with $g_k$ that we do not need explicitly (unless we also want to extract
the amplitudes of the $\Phi_\mu$ correlations, which we do not pursue here) other than $g_k > 0$ from the positive definiteness of $G$. To
incorporate the $r$ dependence, we further define
\begin{subequations}
\begin{eqnarray}
\Lambda(r) & = & {\rm diag}(g_1r^{-2\Delta_1},\ldots,g_Mr^{-2\Delta_M}), \label{lambdadef} \\
S(r) & = & R(r)LR^{-1}(r), \label{smatdef}
\end{eqnarray}
\end{subequations}
so that $\CCp = S(r) \Lambda(r) S^\dagger(r)$. Here $S_{ii}=1$ are inherited from $L$ and the other elements are $S_{ij} = L_{ij}r^{-(\Delta_i - \Delta_j)}$
for $i > j$. Since $S(r)$ is also triangular, $\det\{ S(r)\}=1$; thus, $S(r)$ is invertible and all eigenvalues of $S(r)S^\dagger(r)$ are positive with
product $|\det \{S\}|^2=1$. Their sum is ${\rm Tr}\{SS^\dagger\}=\sum_{ij}|S_{ij}|^2$, which for $r \ge 1$ allows us to bound the largest eigenvalue;
\begin{equation}
\lambda_{\rm max} (SS^\dagger) \le \Sigma_L = \sum_{ij}|L_{ij}|^2,
\end{equation}
independently of $r$. Moreover, since $\prod_i \lambda_i(SS^\dagger)=1$, we have a lower bound
\begin{equation}
\lambda_{\rm min} (SS^\dagger) \ge \Sigma_L^{-(M-1)}.
\end{equation}
Thus, Ostrowski's theorem can be applied as
\begin{equation}
\Sigma_L^{-(M-1)} g_kr^{-2\Delta_k} \le \lambda_k(\CCp) \le \Sigma_L g_k r^{-2\Delta_k},
\end{equation}
which implies $\lambda_k \asymp r^{-2\Delta_k}$, since any other decay power would violate one of bounds for large $r$.
An $m$-fold degeneracy will produce $m$ eigenvalues decaying with the same exponent.

Next, we consider lattice operators, restricting to $Q=0$ in Eq.~(\ref{OPhiexp}) so that positive definiteness can be guaranteed;
treating mixtures with other $Q$ is discussed in the main text. We define the overlap vector $Z_\mu$ for each of the scaling fields
and the full $N\times M$ overlap matrix $Z$
\begin{subequations}
\begin{eqnarray}
Z_\mu & = &(z_{1\mu},z_{2\mu},\ldots,z_{N\mu})^T, \label{Zmudef} \\
Z & = & (Z_1,Z_2,\ldots,Z_M) \label{Zdef}.
\end{eqnarray}
\end{subequations}
The strategy is again to establish bounds for $\lambda_k$.

Using Eq.~(\ref{gkk}) and defining $A=RK$, we can relate the $N\times N$ matrix $\CC$ to the $M\times M$ matrix $\CCp=AA^\dagger$,
\begin{equation}
\CC=Z^*\CCp Z^T=(Z^*A)(A^\dagger Z^T),
\end{equation}
whose nonzero eigenvalues are the same as those of $A^\dagger Z^TZ^*A$. Then, for any vector $c$, with $u=Ac$,
\begin{eqnarray}
&& c^\dagger A^\dagger Z^TZ^*Ac = u^\dagger Z^TZ^*u \nonumber \\
&& ~~~~~~~ \le z_{\max}u^\dagger u= z_{\max}c^\dagger A^\dagger Ac,
\end{eqnarray}
where $z_{\max}$ is the largest eigenvalue of $Z^TZ^*$. Since $A^\dagger A$ has the same eigenvalues as $\CCp=AA^\dagger$,
we can use the ordering theorem (i) to establish an upper bound on the eigenvalues $\lambda_k(r)$ of $\CC(r)$:
\begin{equation}\label{upper}
\lambda_k(r)\le z_{\max}\lambda_k\bigl(\CCp(r)\bigr).
\end{equation}

\begin{figure*}[t]
\includegraphics[width=170mm]{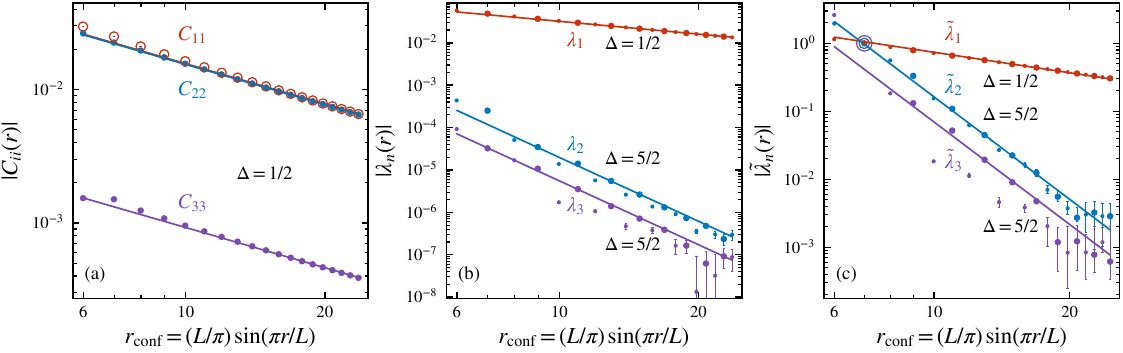}
\caption{Individual correlation functions (a) and covariance eigenvalues (b) computed with the $w=5$ time-reversal even, parity-odd operators
in Table~\ref{otable}. Decay powers corresponding to expected scaling dimensions are indicated by the lines. Smaller and larger circles have
the same meaning as in Fig.~\ref{fig1}. In (c), GEVP eigenvalues with the same operators are shown, using $r_0=7$. Note that these eigenvalues
equal $1$ at $r=r_0$ by definition.}
\label{fig3}
\end{figure*}

The lower bound is more involved but can again be constructed using the ordering theorem (i). For any vector $c$ and $y=Z^Tc$, we have
$c^\dagger \CC c=y^\dagger\CCp y$. Because of the ordering convention, we can split $y$ according to the components $y_l=(y_1,\ldots,y_N)$
belonging to the $N$ lowest fields (the ones dominating for large $r$) and the remaining higher part $y_h=(y_{N+1},\ldots,y_M)$. Then 
\begin{equation}
y^\dagger\CCp y = y_l^\dagger\CCp_{ll}y_l + y_l^\dagger\CCp_{lh}y_h
 + y_h^\dagger\CCp_{hl}y_l + y_h^\dagger\CCp_{hh}y_h,
\end{equation}
and we note that the $l$ and $h$ blocks of $\CCp$ are also individually positive definite and invertible. Since ${\CCp_{hl}}^{\dagger} =\CCp_{lh}$,
by defining $B={\CCp_{hh}}^{-1}\CCp_{hl}$ and $\tilde\CCp=\CCp_{ll}-\CCp_{lh}{\CCp_{hh}}^{-1}\CCp_{hl}$ we can also write
\begin{equation}
y^\dagger\CCp y = y_l^\dagger\tilde\CCp y_l + (y_h+By_l)^\dagger\CCp_{hh}(y_h+By_l).
\end{equation}
Here the second term is nonnegative because $\CCp_{hh}$ is positive definite; thus, for any $c$,
\begin{equation}\label{schur}
c^\dagger \CC c \ge y_l^\dagger\tilde\CCp y_l.
\end{equation}
Because $\CCp=RGR$ and $R$ is diagonal, we have $\tilde\CCp=R_l\tilde G R_l$, where
\begin{equation}
\tilde G=G_{ll}-G_{lh}G_{hh}^{-1}G_{hl},
\end{equation}
and $\tilde G$ is again Hermitian, positive definite, and independent of $r$. Since $y_l=Z_l^Tc$, Eq.~(\ref{schur}) together
with theorems (i) and (ii) gives
\begin{equation}\label{lower}
\lambda_k(r)\ge\lambda_k\bigl(Z_l^*\tilde\CCp(r)\,Z_l^T\bigr)
\ge z_{\min}\,\lambda_k\bigl(\tilde\CCp(r)\bigr),
\end{equation}
where $z_{\min}$ now is the smallest eigenvalue of $Z_l^*Z_l^T$, which is positive if $Z_l$, the block of overlaps with the $N$ lowest fields, is invertible
(i.e., has full rank, which can be generically achieved by a proper choice of lattice operators).
Since $\tilde \CCp$ is also of the simpler form discussed above, its eigenvalues $\lambda_k$, $k=1,\ldots,N$, also decay as $r^{-2\Delta_k}$
and the bounds (\ref{upper}) and (\ref{lower}) imply the final result Eq.~(\ref{lambdan}). The bounds do not, strictly speaking,
guarantee convergence $\lambda_k \to a_k r^{-2\Delta_k}$ with constant $a_k$, e.g., there could be some oscillating correction. This is not
expected on physical grounds, however.

We assumed for simplicity that $G$ in Eq.~(\ref{Cftdef}) is $r$ independent but the proof remains valid as long as $G$ and the overlap matrix $Z$ become
$r$ independent for $r \to \infty$. Lattice models always have corrections from irrelevant operators and in correlation functions (\ref{cdef}) they
are of the general form $[1+\sum_jc_j(\mu,\nu)r^{-\omega_j}]$, with $\omega_j = \Delta^{\rm irr}_j - d - z > 0$, which multiply the asymptotic decay forms
$r^{-(\Delta_\mu+\Delta_\nu)}$. For sufficiently large $r$, these corrections can only propagate in the diagonalization process into similar forms
multiplying the covariance eigenvalues.
\vskip2mm

{\it Appendix B: Further Numerical Results.}---Here
we show results for the covariance matrix based on $O_8$, $O_9$ and $O_{10}$ in Table~\ref{otable}, which use a $5$-site cell with
three simple $p=-1$ operators ${\bf S}_i \cdot {\bf S}_j$ with $i$ and $j$ on different sublattices. As seen in Fig.~\ref{fig3}(a), the
conventional correlators all exhibit decays governed by $\Delta_\epsilon=1/2$. Although the continuum $\epsilon$ field has Lorentz spin
$s=0$ and space reflection parity $+1$, the lattice combination of operators with $p=-1$ compensates the phases of the $\epsilon$
field in Eq.~(\ref{ssop}). Unlike Fig.~\ref{fig1}(a), here we do not see $\Delta_T = \Delta_{JJ}=2$, because these operators are canceled by
$p=-1$ of the lattice operators. Their derivatives with scaling dimension $\Delta'_T = \Delta'_{JJ}=3$ are generated by the operator combinations,
but this is larger than the second $\epsilon$ descendant dimension $\Delta''_\epsilon=5/2$, which is observed in Fig.~\ref{fig3}(b); here the
first descendant operator is also not present, in accord with the selection rule. The doubly degenerate $\Delta''_\epsilon=5/2$ originates from
two second descendants: the one with $s=0$ and the real combination of those with $s=\pm 2$.
\vskip2mm

{\it Appendix C: Generalized Eigenvalues.}---In particle
physics, separation of masses is better achieved with a generalized eigenvalue problem (GEVP) involving a
reference time $\tau_0$ \cite{Lucher90,Alpha09}. This approach was recently taken in real-space \cite{Antunes26}, i.e., for
covariance matrices $\CC(r)$ and a reference distance $r_0$, solving
\begin{equation}
\CC(r) v_k(r, r_0) = \tilde{\lambda}_k(r, r_0) \CC(r_0) v_k(r, r_0),
\label{eq:gevp}
\end{equation}
where $r>r_0$. With $u_k=\CC(r_0)^{1/2}v_k$, Eq.~\eqref{eq:gevp} becomes
\begin{equation}
\begin{gathered}
\widehat{\CC}(r,r_0)\,u_k=\tilde{\lambda}_k(r,r_0)u_k,\\
\widehat{\CC}(r,r_0)=\CC(r_0)^{-1/2}\CC(r)\CC(r_0)^{-1/2}.
\end{gathered}
\end{equation}
Hence the GEVP of matrices $\CC(r)$ and $\CC(r_0)$ is equivalent to the ordinary diagonalization of the matrix $\widehat{\CC}(r, r_0)$.
We here compare the ordinary eigenvalues $\lambda_k$ with $\tilde{\lambda}_k$ obtained with the GEVP.

If the covariance matrix has an idealized fixed-overlap spectral form with $G=1$ in Eq.~(\ref{Cftdef}), then the GEVP would give exactly
$\tilde{\lambda}_k(r, r_0) = (r/r_0)^{-2\Delta_k}$ because GEVP cancels the nonuniversal overlap matrix $Z$ (which depends on the particular
lattice realization and operator basis but not on the scaling dimensions). However, with the realistic nondiagonal form of $G$, the
form is $a(r)(r/r_0)^{-2\Delta_k}$ with $a(r) \not=1$ for $r>r_0$. The potential advantages of the GEVP are not clear.

We present a test with $r_0=7$ in Fig.~\ref{fig3}(c),
where the underlying data are the same as in Fig.~\ref{fig3}(b). There is a change in overall scale but the level of noise is higher
in the GEVP results. We have not found cases where the GEVP clearly outperforms the conventional eigenvalue method, although
typically it is also not much worse.

\end{document}